\documentclass[sigconf,screen]{acmart}

\usepackage{amsmath}
\usepackage{multirow}
\usepackage{xcolor}
\usepackage{array}
\usepackage[shortlabels]{enumitem}
\usepackage{cleveref}

\newcommand{\beginsupplement}{%
        \setcounter{table}{0}
        \renewcommand{\thetable}{S\arabic{table}}%
        \setcounter{figure}{0}
        \renewcommand{\thefigure}{S\arabic{figure}}%
     }

\AtBeginDocument{%
  \providecommand\BibTeX{{%
    \normalfont B\kern-0.5em{\scshape i\kern-0.25em b}\kern-0.8em\TeX}}}

\copyrightyear{2023} 
\acmYear{2023} 
\setcopyright{rightsretained} 
\acmConference[FAccT '23]{2023 ACM Conference on Fairness, Accountability, and Transparency}{June 12--15, 2023}{Chicago, IL, USA}
\acmBooktitle{2023 ACM Conference on Fairness, Accountability, and Transparency (FAccT '23), June 12--15, 2023, Chicago, IL, USA}
\acmDOI{10.1145/3593013.3594099}
\acmISBN{979-8-4007-0192-4/23/06}

\usepackage{etoolbox}
\makeatletter
\patchcmd{\maketitle}{\@copyrightpermission}{
  \begin{minipage}{0.3\columnwidth}
    \href{https://creativecommons.org/licenses/by/4.0/}{\includegraphics[width=0.90\textwidth]{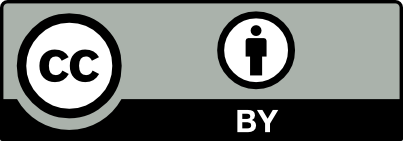}}
  \end{minipage}\hfill
  \begin{minipage}{0.7\columnwidth}
    \href{https://creativecommons.org/licenses/by/4.0/}{This work is licensed under a Creative Commons Attribution International 4.0 License.}
  \end{minipage}

  \vspace{5pt}
}{}{}

\begin{document}

\title{The Progression of Disparities within the Criminal Justice System: Differential Enforcement and Risk Assessment Instruments}


\author{Miri Zilka}
 \affiliation{%
   \institution{University of Cambridge}
   \city{Cambridge}
   \country{UK}}
 \email{mz477@cam.ac.uk}

\author{Riccardo Fogliato}
    \affiliation{%
   \institution{Carnegie Mellon University}
   \city{Pittsburgh}
   \state{Pennsylvania}
   \country{USA}
 }
 
\author{Jiri Hron}
 \affiliation{%
   \institution{University of Cambridge}
   \city{Cambridge}
   \country{UK}}
   
\author{Bradley Butcher}
 \affiliation{%
   \institution{University of Sussex}
   \city{Brighton}
   \country{UK}}

 \author{Carolyn Ashurst}
 \affiliation{%
  \institution{The Alan Turing Institute}
  \city{London}
  \country{UK}
}

\author{Adrian Weller}
 \affiliation{%
   \institution{University of Cambridge}
   \city{Cambridge}
   \country{UK}}
\affiliation{%
  \institution{The Alan Turing Institute}
  \city{London}
  \country{UK}
}

\renewcommand{\shortauthors}{Zilka et al.}

\begin{abstract} \looseness=-1
Algorithmic risk assessment instruments (RAIs) increasingly inform decision-making in criminal justice. 
RAIs largely rely on arrest records as a proxy for underlying crime.
Problematically, the extent to which arrests reflect overall offending can vary with the person's characteristics. 
We examine how the disconnect between crime and arrest rates impacts
RAIs and their evaluation.
Our main contribution is a method for quantifying this bias via estimation of the amount of unobserved offenses associated with particular demographics.
These unobserved offenses are then used to augment real-world arrest records to create part real, part synthetic crime records.
Using this data, we estimate that four currently deployed RAIs assign 0.5--2.8 percentage points higher risk scores to Black individuals than to White individuals with a similar \emph{arrest} record, but the gap grows to 4.5--11.0 percentage points when we match on the semi-synthetic \emph{crime} record.
We conclude by discussing the potential risks around the use of RAIs, highlighting how they may exacerbate existing inequalities if the underlying disparities of the criminal justice system are not taken into account. In light of our findings, we provide recommendations to improve the development and evaluation of such tools.
\end{abstract}

\maketitle

\section{Introduction}

The use of algorithmic decision aid tools is becoming widespread across the criminal justice system, within the US, the UK, and globally \citep{ZilSarWel2022,desmarais2021predictive, mamalian1999use, weisburd2008compstat, babuta2020data, jansen2018data, sprick2019predictive,hamilton2021evaluating,monahan2016risk}. 
These tools assist consequential decisions across the justice system \citep{ZilSarWel2022}, from policing patrol planning \citep{sankin_2021_crime, predpol_how}, deciding which crimes to investigate \citep{EBIT} and whom to charge \citep{HART}, to court pretrial decisions \citep{psa, van2019use, OGRS3} and prison security classification \citep{ZilSarWel2022}. One prominent category of tools are risk assessment instruments (RAIs), i.e., tools that predict the risk an individual will re-offend in the future. 

\looseness=-1
Historically, the justice system relied on experts---mainly mental health professionals---to evaluate the risk posed by an individual \citep{goeShrSke2021}. 
However, some research suggests that statistical predictions outperform expert recommendations given the same information \citep{Meehl1954,DawFauMee1989,AegWhiSpe2006}. 
This has been shown both for criminal behavior and violence predictions made by judges \citep{Gottfredson1999,Krauss2004,JunConShr2017}, and by probation officers \citep{CohPenVan2016}. 
However, there is also a body of work demonstrating
the risks posed by biased algorithmic systems \citep{barocas2016big,sankin_2021_crime, chouldechova2017fair, flores2016false}. 
Scholars have warned of the risks of using data that encodes inequalities \citep{ensign2018runaway, lum2016predict, richardson2019dirty},
which may be exacerbated by harmful feedback loops \citep{lum2016predict, ensign2018runaway}, and lack of transparency \citep{winston2018palantir}.

\looseness=-1
While algorithmic systems
deployed within the criminal justice system are varied \citep{fitzpatrick2019keeping, berk2013statistical, fitzpatrick2020policing, hunt2014evaluation, mohler2015randomized}, 
their common feature is reliance on criminal activity records. 
Problematically, \emph{observed} crime is a poor proxy for \emph{overall} crime \citep{brame2004criminal, Fogliato_2021}. 
Disparities in observed crime rates can be a result of differences in witness reporting or in discovery (e.g., patrolling patterns), and 
the likelihood a crime will become known to the law enforcement can vary significantly depending on a person’s sensitive attributes, such as sex, race, and age \citep{ButRobZil2022, bosick2012reporting, baumer2010reporting, fitzpatrick2019keeping}. 
Works including those by \citet{akpinar2021effect, ensign2018runaway}, and \citet{lum2016predict} have explored the impact of differential reporting and discovery rates on \emph{predictive policing}, showing how these can lead to outcome disparities and feedback loops.

\looseness=-1
In contrast to this prior work, we focus on \emph{risk assessment instruments} (RAIs).
While risk prediction is the nominal goal, RAIs are trained 
and evaluated on their ability 
to predict the risk of re-arrest.
Evaluation using arrest data is problematic.
In the presence of subpopulation arrest rate disparities,
a RAI that predicts risk arrest accurately may appear superficially unbiased, 
and yet contain implicit biases by not adjusting for the disproportionately high chance of re-arrest in certain subpopulations.
In this work, we undertake a \emph{quantitative} investigation of the impact of differential reporting and discovery rates on risk assessment scores. 
While \emph{theoretical} explorations have been presented by others \citep{eckhouse2019layers,goel2021accuracy}, this is to the best of our knowledge the first attempt to quantify the magnitude of the potential effect.

\looseness=-1
We begin by discussing how RAIs are used within the criminal justice system, and how these impact individuals who interact with the system (\Cref{sec:background}). 
In \Cref{sect:observed_comparison}, we use data from Harris County, the Texas section of the Neulaw Criminal Record Database \citep{ormachea2015new}, 
to study the average difference in RAI scores between individuals matched on arrests, 
without taking their unobserved criminality into account.
In \Cref{sect:arrests_bias}, we describe how to use the US National Survey on Drug Use and Health (NSDUH), and the National Crime Victimization Survey (NCVS), to simulate the unobserved crimes associated with individuals from the Neulaw cohort.
In \Cref{sect:synth_comparison}, we combine the simulated unobserved crime with observed arrests to run analysis analogous to \Cref{sect:observed_comparison} except for matching individuals on their semi-synthetic crime history instead of arrests.
We provide a sensitivity analysis in \Cref{sec:results_sensitivity}. 
In \Cref{sec:discussion}, we conclude by describing the limitations and ethical considerations of our work, and provide a set of recommendations based on our results.


Our main contributions are:
\begin{enumerate}[topsep=1pt]
    \item \looseness=-1
    We propose a method for quantifying implicit bias within RAIs due to differential arrest rates. 
    
    \item \looseness=-1
    We use semi-synthetic data to empirically investigate the degree to which four currently deployed RAIs are impacted by differential arrest rates.
  
    \item \looseness=-1 
    We find evidence that all investigated RAIs are impacted by differential arrest rates, with 
    Black individuals 
    disproportionately affected.
    The estimated difference in RAI scores is 0.5--2.8 percentage points when matching only on arrests (\Cref{tab:baseline}), but rises to 4.5--11 percentage points when unobserved crime is taken into account (\Cref{tab:main_results}).
\end{enumerate}
In response to our findings, in \Cref{sect:conclusion} we provide recommendations regarding responsible design and evaluation of RAIs, and the data required to better quantify and mitigate the effects we observe.  

\section{Background}
\label{sec:background}

\subsection{The criminal justice pipeline}

\begin{figure}[t]
    \centering
    \includegraphics[width=\linewidth]{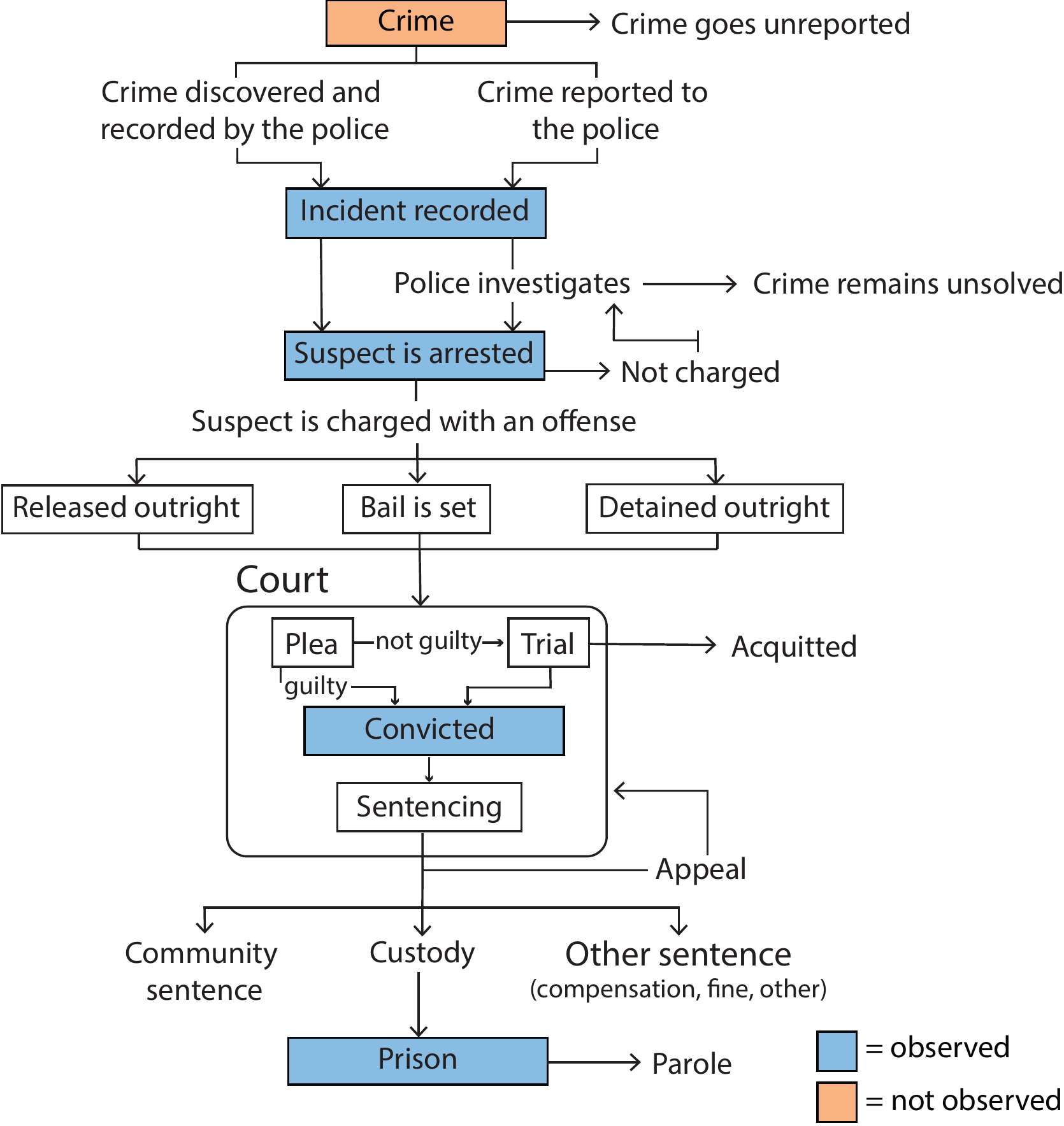}
    \caption{An illustration of the criminal justice pipeline and the relationships between crime, recorded incidents, arrests, convictions, and prison sentences respectively. This diagram is a simplification, and does not specify all possible pre-trial and sentencing outcomes (see e.g., \citep{CJ_flowchart}, for a more detailed diagram).}
    \label{fig:pipeline}
    \vspace{-1.5em}
\end{figure}

\looseness=-1
We conceptualize the law enforcement as a funnel-shaped pipeline (\Cref{fig:pipeline}), 
beginning with all criminal activity on top, and sentencing and incarceration at the bottom. We highlight that this diagram is for illustrative purposes only, and does not specify all possible pre-trial and sentencing outcomes (see e.g., \citep{CJ_flowchart}, for a more detailed diagram, and \citep{mayeux2018idea,neubauer2018america} for a more comprehensive description of the justice system).
Here we describe the pipeline, to show how RAIs fit within an individual's wider experience of the justice system. 

\looseness=-1
Starting at the top, not all illegal activity becomes known to law enforcement agencies. 
We consider two ways a criminal offense may become known: 
(i)~a report to the police, either by a victim or by some third party, or
(ii)~a discovery through proactive policing efforts. 
The latter applies to offenses such as driving under the influence (DUI),
and possession of illegal substances. 
If a crime is discovered by law enforcement, 
an officer may arrest the person who committed an offense, issue a fine or a citation, or let the person go with an informal warning. 
In all cases except the latter, the crime is usually, but not always, recorded by the law enforcement agency. We note that incidents reported to the police might also not be recorded as a crime, or lead to no further investigation. 

Irrespective of how the crime became known to law enforcement, 
after a suspect has been identified the agency will decide whether they should be charged. 
The first decision after a person has been charged is their pre-trial custody status. A suspect can be released outright, released conditional on non-monetary conditions, on bail, or detained outright. 
Charges will then turn into convictions only if the court finds the suspect guilty and decides to issue a conviction. 
If the suspect is found guilty, a sentence will be decided by the judge. 
Generally, sentencing  guidelines set minimum and maximum terms for the sentence, including limits on the time of imprisonment and the amount that can be set for monetary fines. We note that Sentencing guidelines can be advisory (e.g.,  Federal Sentencing Guidelines in the US) or mandatory (e.g., in the UK ``the courts must follow any relevant sentencing guidelines, unless it is contrary to the interests of justice to do so''). 
The minimum sentence may depend on the individual's prior convictions. 
In some cases, probation may be an alternative to incarceration. 
If the sentence does include an incarceration period, once the minimum period of incarceration has elapsed, the individual may be granted parole and released before serving their full sentence. For long sentences, a decision about whether to grant parole may take place periodically.

\subsection{Examples of RAIs in criminal justice}



\looseness=-1
Algorithmic risk assessment instruments are commonly used as aids at pre-trial and sentencing across most US states \citep{desmarais2021predictive, stevenson2021algorithmic, fazel2021predictive}. These RAIs often target the following three outcomes: 
(i) the risk that the individual fails to appear in court at trial; 
(ii) the risk of re-arrest in a time bounded period, e.g., within two years from the time of release from prison; 
and (iii) the risk of re-arrest for a violent offense within a certain time window.
RAIs are also used to determine the level of threat posed by incarcerated people to their peers and prison officers,
\citep{massaro2021analyzing}, in probation and parole decisions \citep{oasys}, and to assess risk of violence \citep{singh2014international}.

\looseness=-1
\emph{Most RAIs make use of individual's prior arrest or conviction data.}
While simulation experiments demonstrate that RAI predictions may be more accurate than those of judges \citep{Kleinberg2017},
\citet{stevenson2021algorithmic} show that adoption of a RAI by judges in Virginia led to no change in the incarceration and crime rates, and no evidence of change in sentencing racial disparities. 
It is unknown to what extent the judges paid attention to the RAI's recommendations. 

\looseness=-1
Scholars from multiple disciplines have questioned the use of RAIs. 
One of the primary concerns is bias \citep[e.g.,][]{mayson2019bias, hamilton2019sexist,huq2018racial, berk2023fair}, particularly racial \citep[e.g.,][]{brennan2009evaluating,angwin2016machine}.
Other works have cast doubt on RAI's scientific validity \citep{eaglin2017constructing,hamilton2020judicial}, and highlighted legal and ethical issues \citep{monahan2016risk,hamilton2015risk}. 
Notably, the validity of using arrests as a proxy for actual offences has been questioned, as well as the fact that 
RAIs can only be evaluated using data from \emph{released} individuals \citep{Fogliato_2021,eckhouse2019layers,goel2021accuracy,bao2021s}.

\looseness=-1
\section{RAI comparison based on arrests}
\label{sect:observed_comparison}

\looseness=-1
In an ideal world, RAIs should assign the same risk level to individuals with a similar risk of re-offending. 
In practice, not all offenses are observed, necessitating the use of proxies.
All RAIs we consider here compute scores using a combination of arrest history and demographic attributes.
Race, ethnicity, and gender are not explicitly used as input variables, in an attempt to rid RAIs of biases \citep{SimAdaWel2021}. 
However,
simply omitting protected attributes from algorithm inputs may not eliminate bias \citep{barocas2016big}. 
We therefore examine the relationship between RAI predictions and race and ethnicity.

\looseness=-1
Our goal is to estimate the average difference in RAI scores between individuals who differ in race or ethnicity, but are matched based on age, gender, and criminal history.
The next sections describe the offense data (\Cref{sect:creating_cohort}), the RAIs (\Cref{sect:calculating_rais}), and the matching algorithm (\Cref{sect:estimating_effect}) we use to estimate the effect.
In \Cref{sect:baseline_results}, we present the baseline effect estimates, which---similarly to the RAIs---use arrests in place of criminal history.
This will lay the basis for the next sections, where we examine the bias introduced into RAI \emph{evaluation} by treating arrests as a proxy for crime.

\subsection{Collating a cohort of individuals with available arrest record}
\label{sect:creating_cohort}

We use the Harris County, Texas arrest records retrieved from the Neulaw criminal record database \citep{zilka2022a,ormachea2015new}. 
This dataset consists of 3.1 million records from the Harris County District Clerk's Office, spanning from 1977 to April 2012.
Due to missing records in earlier years, we only consider 1992--2012 records. 
The data includes 39 variables including the type of crime, the date of birth, sex and race of the person who was charged with the offense, and various criminal case details. 
We chose this dataset because it contains unique individual identifiers, enabling us to construct a criminal record for each person.\footnote{We note that these records may not be complete as the dataset does not contain information on records outside this jurisdiction.}

\subsection{Calculating risk scores}
\label{sect:calculating_rais}

For each individual, we calculate four different risk assessment instrument scores: 
\begin{enumerate}[(a),leftmargin=1.75em]
    \item 
    the Public Safety Assessment's New Criminal Arrest (NCA);
    
    \item
    the Public Safety Assessment's New Violent Criminal Arrest (NVCA);\footnote{The first part of the PSA is the Failure to Appear (FTA) score. We are not able to produce reliable estimates for the FTA score using the variables in the HC dataset and so it is not included in our analysis.} 
    
    \item
    the Virginia Pretrial Risk Assessment Instrument (VPRAI) \citep{VPRAI}; 
    
    \item
    the Offender Group Reconviction Scale version 3 (OGRS3) \citep{OGRS3}. 
\end{enumerate}
All risk scores are estimated based on the public description of the tool's algorithm and the information contained within the dataset. See Appendix~\ref{Apx:scores} for more details.

\subsection{Matching similar individuals}
\label{sect:estimating_effect}

We want to match individuals with similar offense history, age, and sex,\footnote{Unfortunately, none of the datasets records gender identity.} but different race or ethnicity.
We then compute the average difference in the RAI assigned risk scores of the matched individuals.
For each RAI we measure two metrics: 
\begin{enumerate}[leftmargin=1.75em]
    \item 
    the \emph{Conditional Average Effect} (CAE), which is the average difference in score for a specific group of matched individuals (e.g., White men, aged 18-30, who committed a single aggravated assault will be matched with Black men, aged 18-30, who committed a single aggravated assault);
    
    \item 
    the \emph{Average Effect} (AE), which is the weighted average of the effect over all matched groups.
\end{enumerate} \looseness=-1
We use the \texttt{DAME-FLAME} Python library to obtain the matching \citep{gupta2021dame,liu2018interpretable,wang2021flame}.
Instead of feeding in age and number of arrests for each offense type as integers, we bin both into discrete categories.\footnote{We match based on the number of arrests per offense type (e.g., property, DUI), not the overall number of arrests.}
This ensures that people close in age and with similar arrest counts for similar offences get matched.
For age, we use fixed binning into categories \{18--29, 30+\}.
For arrest counts, we use \{0, 1, 2, 3--4, 5--6, 7--9, 10--19, 20--49, 50+\}.
In the sensitivity analysis (\Cref{sec:results_sensitivity}), we investigate the impact of different binning choices.

\begin{table*}[tb]
\centering
\caption{Average percentage point difference in NCA, NVCA, VPRAI and OGRS3 scores between matched individuals}
\label{tab:baseline}
\begin{tabular}{lcccc}
\hline
Compared populations & NCA & NVCA & OGRS3 & VPRAI \\
\hline
Black--White & 
$1.11\% \pm0.25\%$ &
$0.88\% \pm0.24\%$ &
$0.54\% \pm0.27\%$ &
$2.76\% \pm0.46\%$ \\
Hispanic--White (non-Hispanic) & 
$0.75\% \pm0.33\%$ &
$0.30\% \pm0.24\%$ &
$1.19\% \pm0.23\%$ &
$0.58\% \pm0.19\%$ \\
\hline

\multicolumn{5}{c}{\parbox[t][][t]{0.8\textwidth}{\vspace{0.3em}\footnotesize{Average Effect (AE) are the average difference in risk scores between matched groups of Black and White individuals (first row) or Hispanic and White (non-Hispanic) individuals (second row), when matching on age, sex, and arrest records. Both presented as a percentage of the score range of the respective RAI. Errors are calculated as the standard deviation of five runs with different seeds. 
Results are matched on the largest set of crime bins, i.e., $\{0,1,2,3,4-5,5-6,7-9,10-19,20-49,50+\}$.}}}
\end{tabular}
\end{table*}

\subsection{Baseline results}
\label{sect:baseline_results}

\looseness=-1
For comparison, we estimate the differences in risk scores for matched individuals (without accounting for differences in reporting/discovery rates); see \Cref{tab:baseline}. 
We use arrest data from 10,000 randomly sampled individuals from the HC dataset. We report the average difference in risk scores between matched groups of Black and White individuals (first row) and Hispanic and White individuals (second row).
Matching is done over age, sex, and arrest records. 
Different RAIs use different score ranges;
within our cohort, NCA and NVCA scores are between 1 and 6, OGRS3 is between 0.006 and 0.96, and VPRAI is between 0 and 5.
To be able to compare the magnitude of the effect, we normalize all scores to the $[0, 1]$ interval, and present the results as a percentage point change with respect to these. 
For example, the table shows the average difference in NCA scores between Black and White individuals is 1.11 percentage points. 
This translates to an average difference in (non-normalised) score of $0.0111 \times 5 = 0.0555,$ since NCA has a range of 5.
With the exception of VPRAI with respect to Black individuals, all AEs were around or under $1\%$ of the score range. 

\section{Why can a comparison based only on arrest distort measured bias?}
\label{sect:arrests_bias}

\begin{figure*}[t]
    \centering
    \includegraphics[width=0.72\linewidth]{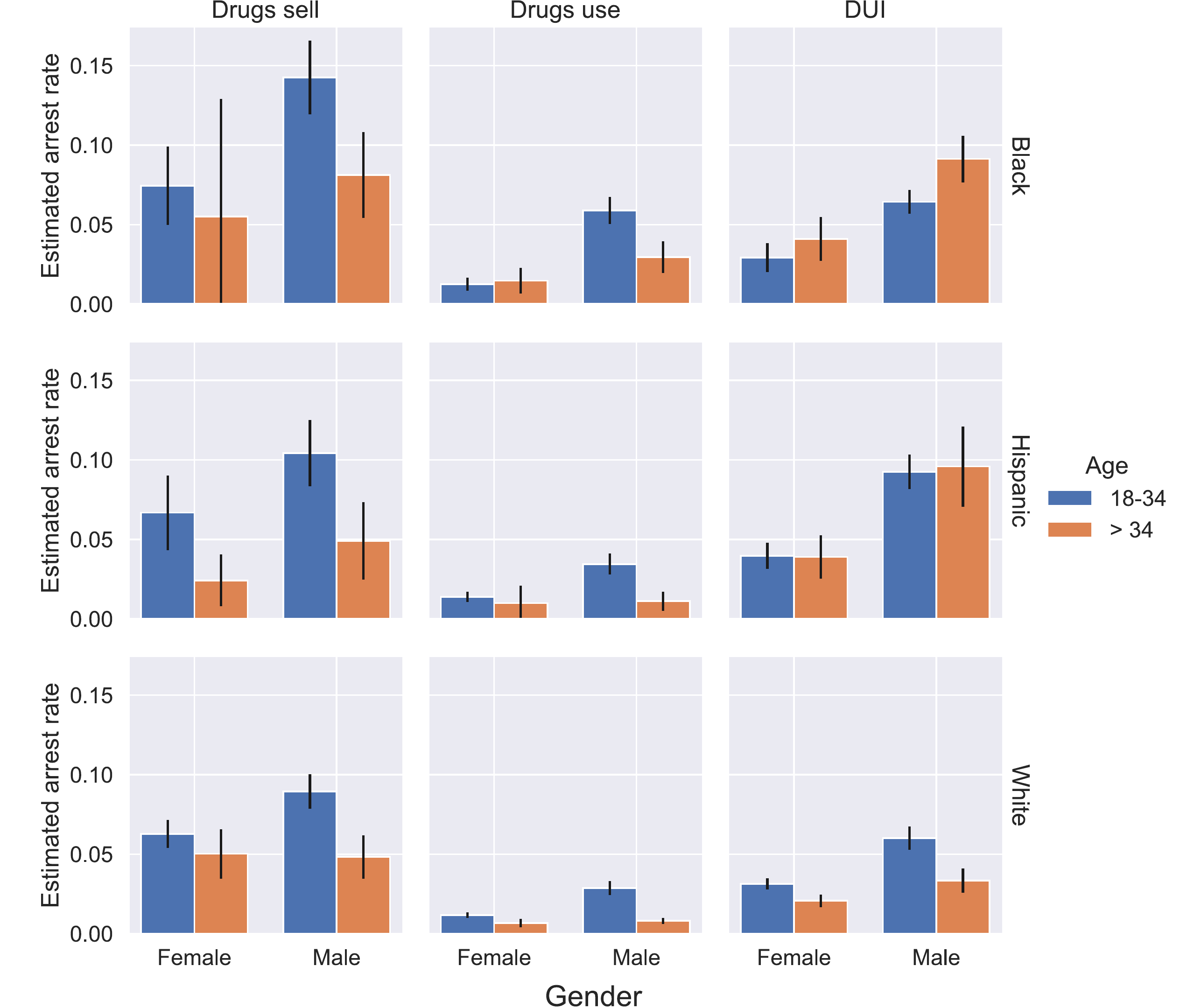}
    \caption{Estimates of arrest rates and corresponding standard errors by sex, age and race of people who committed drug use, drug sale, and DUI offenses. Rates are estimated from the National Survey on Drug Use and Health (NSDUH) as the average arrest rate for respondents that self-reported engaging in these illegal activities in the year prior to the interview. For DUI and drug offenses, Black and Hispanic males more likely to be arrested for their crime compared to White males.}
    \label{fig:arrest_rates_nsduh}
\end{figure*} 

\begin{figure*}[t]
    \centering
    \includegraphics[width=\linewidth]{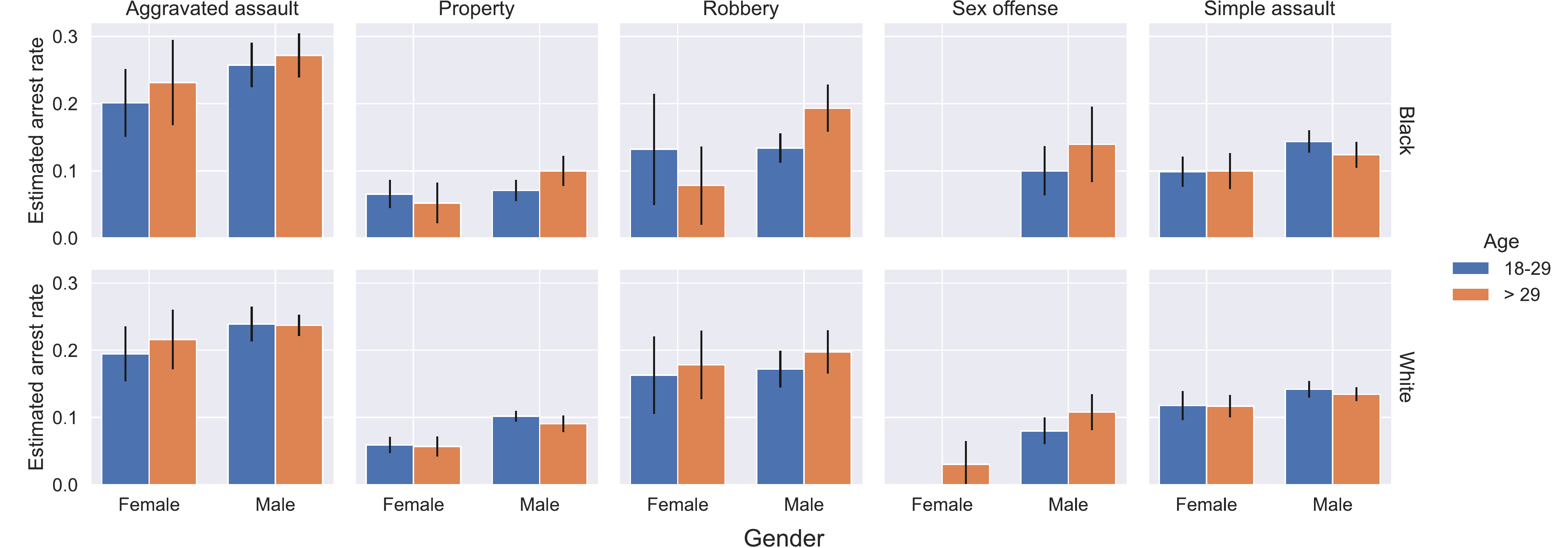}
    \caption{Estimates of arrest rates and corresponding standard errors by sex, age and race of people who committed property and violent offending. Rates are estimated from the National Crime Victimization Survey (NCVS) as the average arrest rates for assaults, robbery, property and sexual offenses conditional on the demographics of the person who committed the offense.}
    \label{fig:arrest_rates_ncvs}
\end{figure*}







The results in \Cref{sect:baseline_results} suggest that bias in RAIs is relatively modest.
There is a catch though: we matched people based on their arrests rather than on the offenses they committed.
This is problematic because arrest rates---given a crime was committed---are \emph{not} equal across subpopulations \citep{Fogliato_2021,ButRobZil2022}, as apparent from the data (\Cref{fig:arrest_rates_nsduh,fig:arrest_rates_ncvs}).
A higher arrest rate means a higher probability of re-arrest, but not necessarily a higher risk of re-offending.
Matching on arrest history, as in \Cref{sect:observed_comparison}, may thus result in comparing people with rather different crime propensity.
The RAIs can therefore seem less biased, because the \emph{evaluation} itself is predicated on the incorrect assumption of equal arrest rates.

\looseness=-1
This brings us to our main question:
Can we estimate racial bias in RAIs for people with similar offense rather than arrest history?
To answer this, we face the problem of estimating underlying offending.
While the task is essentially impossible at the level of individuals, 
our analysis approach---see \Cref{sect:observed_comparison}---only requires matching crime levels at the subpopulation level.
These crime rates can be estimated from NSDUH and NCVS data.
We therefore utilize NSDUH and NCVS to estimate the number of offenses which did not result in arrest, and distribute a proportion of these to the population analyzed in \Cref{sect:observed_comparison}.
Generation of the unobserved offenses is impossible without further assumptions.
The aim of the following paragraphs is therefore dual:
(a)~describing our approach, and
(b)~illustrating the effects ignored in RAI development and evaluation when arrests are naively treated as a proxy for crime.

\looseness=-1
Our strategy starts by estimating the number of offenses.
For each offense type $c$ and subpopulation $g$ (based on age, gender, and race), the total offense count $N_g^c$ can be estimated using the subpopulation arrest count $A_g^c$ and rate $AR_g^c$
\begin{align}
\label{eq:total_crimes}
    \text{crimes}
    =
    \frac{\text{arrests}}{\text{arrest rate}}
    \, ,
    \quad
    \text{i.e.,}
    \quad
    N_g^c
    =
    \frac{A_g^c}{AR_g^c}
    \, .
\end{align}
The number of unobserved crimes 
is then simply the difference between the total crimes and the arrests $N_g^c - A_g^c$.
However, we need unobserved offenses corresponding to the population of convicted individuals from \Cref{sect:observed_comparison}.
Since some of the $N_g^c$ crimes were committed by people outside of the cohort, i.e., those without an arrest record, the unobserved total for our cohort is 
\begin{align}
\label{eq:unobserved_crime}
    U_g^c \coloneqq \lambda_g^c \cdot N_g^c - A_g^c \, ,
\end{align}
where $\lambda_g^c$ is the re-offense rate among people with a prior arrest.\footnote{We ensure $\text{AR}_g^c \leq \lambda_g^c \leq 1$ to avoid a negative number of unobserved crimes $U_g^c$.}
$U_g^c$ crimes of type $c$ are then distributed among subpopulation $g$.

\looseness=-1
The overall approach we use to generate crime records from arrests can be summarized as follows:\footnote{Code for all experiments, including the assignment of unobserved crimes, is available at \url{https://github.com/Bradley-Butcher/pipeline_paper_code}.}
\begin{enumerate}[leftmargin=1.75em]
    \item 
    Take the Harris County cohort of people with an existing arrest record \Cref{sect:creating_cohort}.
    
    \item 
    Use NCVS and NSDUH national surveys to estimate arrest rates for each crime and subpopulation (\Cref{sect:arrest_rate_estim}).

    \item 
    Use the arrest rates to estimate the number of unobserved crime in each cohort subpopulation (\Cref{sect:lambda_estim}).
    
    \item \looseness=-1
    Distribute the unobserved crimes among the HC cohort based on their history and demographics (\Cref{sect:crime_assignment}).
\end{enumerate}

\subsection{Estimating arrest rates $\text{AR}_g^c$}
\label{sect:arrest_rate_estim}

To estimate arrest rates, we require both data on baseline offending and subsequent arrests. 
We use two datasets:
\begin{itemize}[leftmargin=1.75em]
    \item 
    \textbf{NSDUH}, 
    years 1992--2019,
    for drug and driving under the influence (DUI) offenses. 
    
    \item 
    \textbf{NCVS}, 
    years 1992--2020,
    for property offences,\footnote{We only included property offenses when the demographics of the person who committed the offense were known.} simple assault, aggravated assault, robbery, and sex offenses. 
\end{itemize} 

For \emph{NSDUH}, we estimate the arrest rate for respondents that self-reported engaging in the relevant illegal activity in the year prior to the interview. \footnote{When an individual reported selling drugs, we only included them in the the `drug sell' category, even if they also reported using drugs. 
Thus, the arrest rate for `drug use' only include those individuals that did not engage in drug selling activities.} 
For \emph{NCVS}, we compute the average arrest rates based on individual victim reports, and the proportion of subsequent arrests. 
We note that many crimes will go unreported.
It is thus likely we underestimate the number of unobserved crimes.
Furthermore, both NSDUH and NCVS are national surveys, whereas our cohort is constrained to Harris County, Texas.
While county-wise differences may be significant,
NCVS and NSDUH are to the best of our knowledge the most suitable open data source for US arrest rate estimates \citep{zilka2022a}.\footnote{Other sources---e.g., \citep{beck2021race}---do not provide arrest rates grouped by gender, age, and race, which makes them unsuitable for our matching analysis. 
They also often miss DUI and drug offense data, which can have outsized impact on RAIs due to their high number and the arrest rate discrepancies (see \Cref{fig:arrest_rates_nsduh}).}

\looseness=-1
Since we are estimating arrest rates for every year between 1992 and 2012, we do not have enough data to produce reliable numbers for certain subgroups and crime types.
We thus use the following methods to regularize our estimates:
\begin{itemize}[leftmargin=1.75em]
    \item 
    \textbf{Averaging.}
    Set the arrest rate to the subpopulation average over \emph{all} years, weighting by the number of corresponding entries in NSDUH (resp.\ NCVS).
    
    \item
    \textbf{Regression.}
    Use linear regression to regress arrest rates on year, weighting by the number of corresponding entries in NSDUH (resp.\ NCVS).
    Set the arrest rate to the model predictions clipped to $[0, 1]$.
\end{itemize}

\looseness=-1
Finally, both NSDUH and NCVS contain information about \emph{Hispanic ethnicity}. 
However, the sample size from NCVS is too small to get reliable arrest rate estimates for the Hispanic subgroups. 
As a result, when using arrest rates from NSDUH, we distinguish between White, Black and Hispanic individuals, while for NCVS rates, we use Black and White only. 
This is possible as race and ethnicity are recorded in different columns.\footnote{In Neulaw dataset, Hispanic ethnicity is estimated from the name.}
Hence, for an individual with `White' in the race column, and `Hispanic' in the ethnicity column, drug and DUI offenses (NSDUH data) are assigned based on the relevant Hispanic arrest rates.
All other offenses (NCVS data) will be generated using the relevant White Hispanic arrest rates.

\subsection{Modeling re-offense rates $\lambda_g^c$}
\label{sect:lambda_estim}

\looseness=-1
The parameter $\lambda_g^c$ controls the proportion of total crime assigned to individuals in our cohort, i.e., people with at least one arrest within the 20-year window covered by the Neulaw dataset.
We take two approaches to setting the $\lambda_g^c$ value.
The former first $\lambda_g^c$ as a \emph{fixed} parameter, equal across all groups, and varies it as part of the sensitivity analysis (\Cref{sec:results_sensitivity}).
The second estimates $\lambda_g^c$ from NSDUH for drug and DUI offenses, and from Neulaw for other offenses.

For \emph{NSDUH} (drugs and DUI), we take
\begin{align*}
    \lambda_g^c 
    =
    \frac{
        \bigl| \, g_\text{report}^c \cap g_\text{arrest} \, \bigr|
    }{
        \left|\, g_\text{arrest} \,\right|
    }
    \, ,
\end{align*} \looseness=-1
where $g_\text{report}^c$ are the individuals from group $g$ who self-reported committing a crime of type $c$ within the last year, and $g_\text{arrest}$ are the individuals in $g$ that have at least one arrest (for any crime and year).
The numerator is thus the number of individuals committed a crime of type $c$ in the last year \emph{and} have ever been arrested for \emph{any} crime.

As NCVS surveys victims instead of those who committed the offence, we cannot perform the above calculation for crimes not covered by NSDUH.
Instead, we use \emph{Neulaw} to compute the total number of arrests for each individual, and approximate $\lambda_g^c$ by the proportion of individuals with two or more total arrests for crime $c$ in group $g$.
In other words, we approximate the rate of re-offense by the rate of re-arrest in these cases.

\subsection{Distributing the unobserved crimes $U_g^c$}
\label{sect:crime_assignment}

Using the estimated $\lambda_g^c$ and $\text{AR}_g^c$, we can compute $U_g^c$ using \Cref{eq:total_crimes,eq:unobserved_crime}.
$U_g^c$ represents the number of unobserved crimes of type $c$ committed by individuals within the Neulaw cohort which belong to the demographic group $g$.
We thus need to distribute the $U_g^c$ crimes among the members of group $g$.

This allocation can be done in many ways. 
Since we only have limited information, we choose a stochastic scheme.
In particular, for each crime type $c$, we sample $U_g^c$ individuals with replacement from the group $g$.
Naively, we could sample uniformly at random. 
However, this does not adjust for the increased probability of committing the same crime several times
(e.g., a person with theft arrest is more likely to have committed more thefts than a person with only DUI arrests).
Each individual $i$ in group $g$ is therefore assigned a crime dependent probability $P_i^c$, and the $U_g^c$ samples are then drawn i.i.d.\ from the Categorical distribution with parameter $P_g^c \coloneqq  [ P_i^c ]_{i \in g}$.
We choose 
\begin{align}
\label{eq:crime_weight}
    P_i^c
    \propto
        \tfrac{1}{| g |}U_g^c
    +
    \omega \cdot
        A_{i}^c
    \, ,
\end{align}
where 
$A_i^c$ is the number of arrests for crime $c$ of individual $i$, 
$U_g^c / | g |$ is the number of unobserved crimes $c$ per person in $g$, 
and $\omega \geq 0$ a parameter which determines how prior arrests affect re-offending (without arrest).
We chose to include the $U_g^c / | g |$ in \Cref{eq:crime_weight} to adjust for the disparate influence of prior arrests for different crime categories.
For example, drug use is associated with very low arrest rates (\Cref{fig:arrest_rates_nsduh}), which means that the number of unobserved crimes $U_g^c$ is typically high (see \Cref{eq:unobserved_crime}).
This increases $U_g^c / | g |$, making prior arrest $A_{i}^c$ less influential in determining the probability of further offenses given a fixed $\omega$.

\begin{table*}[]
\centering
\caption{Average percentage point difference in NCA, NVCA, VPRAI and OGRS3 scores between matched Black and White, and Hispanic and White (non-Hispanic) individuals.}
\label{tab:main_results}
\begin{tabular}{lcccc}
\midrule
\multicolumn{5}{c}{Black--White} \\
\midrule
& NCA & NVCA & OGRS3 & VPRAI  \\
\midrule
Simulated Average Effect & 
$7.12\%$ $\pm0.25\%$ &
$4.65\%$ $\pm0.06\%$ & 
$4.47\%$ $\pm0.16\%$ &
$10.72\%$ $\pm0.46\%$ \\
Difference in effect & 
$6.01\%$ $\pm0.35\%$ &
$3.77\%$ $\pm0.25\%$ & 
$3.93\%$ $\pm0.31\%$ &
$7.96\%$ $\pm0.65\%$ \\ 
Proportional increase in effect & 
$542\%$ $\pm258\%$ &
$429\%$ $\pm224\%$ & 
$727\%$ $\pm515\%$ &
$289\%$ $\pm118\%$ \\ 
\midrule
\multicolumn{5}{c}{Hispanic--White (non-Hispanic)} \\
\midrule
& NCA & NVCA & VPRAI & OGRS3 \\
\midrule
Simulated Average Effect & 
$0.94\%$ $\pm0.33\%$ &
$0.28\%$ $\pm0.21\%$ & 
$0.89\%$ $\pm0.13\%$ &
$0.19\%$ $\pm0.10\%$ \\ 
Difference in effect & 
$0.19\%$ $\pm0.47\%$ &
$-0.02\%$ $\pm0.32\%$ & 
$-0.30\%$ $\pm0.27\%$ &
$-0.39\%$ $\pm0.19\%$ \\ 
Proportional increase in effect & 
$26\%$ $\pm19\%$ &
$-8\%$ $\pm10\%$ & 
$-25\%$ $\pm12\%$ &
$-68\%$ $\pm50\%$ \\ 
\hline
\multicolumn{5}{l}{\parbox[t][][t]{0.8\textwidth}{\vspace{0.3em}\footnotesize{Baseline and Simulated Average Effect (AE) are the average difference in risk scores between matched groups of Black and White individuals, when matching on age, sex, and simulated criminal history and arrest records. Both presented as a percentage of the score range of the respective RAI. Difference in effect is the simulated AE minus the baseline AE. Proportional increase is the  difference normalized by the baseline AE, in percentage. 
Results are presented for a simulation with $\lambda=1$, $\omega=1$, $\Delta T=10$ and matched on the largest set of crime bins, i.e., \{0, 1, 2, 3, 4--5, 5--6, 7--9, 10--19, 20--49, 50+\}.}}}
\end{tabular}
\end{table*}

\subsubsection{Choosing $\omega$}
\label{sec:omega}

\looseness=-1
Low $\omega$ makes sampling close to uniformly random, whereas $\omega \gg 1$ translates into assuming that arrests and unobserved crimes are distributed equally within the given subpopulation.
To the best of our knowledge, there is no publicly available data which would allow estimation of $\omega$.\footnote{This will require longitudinal studies on offending on a large adult population.} 
We will therefore experiment with various choices of $\omega$ in the sensitivity analysis (\Cref{sec:results_sensitivity}).

\subsubsection{Time window}
\label{sec:time_window}
To prevent an adverse impact on individuals who were not re-arrested for long periods of time---e.g., a 40-year old with a single arrest from age 18---we compute $P_g^c$ only using data from a rolling time window of size $\Delta T > 1$.
In each window, the resulting probabilities are used to generate unobserved crimes for the last year of the window.
To avoid overcounting, we divide the number of offenses to be generated $U_g^c$ by the window size $\Delta T$;
the only exception is the first window, where we sample for all years in the interval.
For example, when $\Delta T = 5$, the assignment will first be done for years 1992--1997, and then for 1998 based on years 1993--1998, for 1999 based on years 1994--1998, and so on.
This ensures that only arrest records within 5-year window will affect the individual crime assignments. 
We investigate the impact of $\Delta T$ in the sensitivity analysis (\Cref{sec:results_sensitivity}).

\section{RAI comparison based on crime}
\label{sect:synth_comparison}

\looseness=-1
As in \Cref{sect:baseline_results}, 
we sample 10,000 individuals from the Neulaw dataset (all over 18, with no missing demographic data), and compute the RAI scores for each one.
We generate unobserved offenses for each sampled individual as described in \Cref{sect:arrests_bias}, and match on age, sex, and the total number of crimes rather than just arrests.
We then compare scores between matched groups of 
1)~Black and White individuals, and 
2)~Hispanic and White (non-Hispanic) individuals. 

\looseness=-1
The results are in \Cref{tab:main_results}. 
Compared to the arrests-only results (\Cref{tab:baseline}),
the magnitude of disparities between the Black and White subpopulations increases significantly. 
We do not observe the same effect for the Hispanic subpopulation, despite having higher arrest rates than their White (non-Hispanic) counterparts for several crime types (\Cref{fig:arrest_rates_nsduh}).
This may be due to data limitations, as NCVS does not contain data for the Hispanic subpopulation, which means we instead take the arrest rates for White individuals (see \Cref{sect:arrest_rate_estim}).
The arrest rates used to generate unobserved offenses (\Cref{sect:crime_assignment}) thus only differ for drug and DUI related offenses (NSDUH data).
Another potential explanation is that our arrest rate data come from the national NCVS and NSDUH surveys (\Cref{sect:arrest_rate_estim}), whereas the population is constrained to the Harris County, Texas which has much larger than average Hispanic population.
The true arrest rates can thus be significantly different (see \Cref{sec:ethics} for further discussion).

\begin{figure*}[t]
    \centering
    \includegraphics[width=\linewidth]{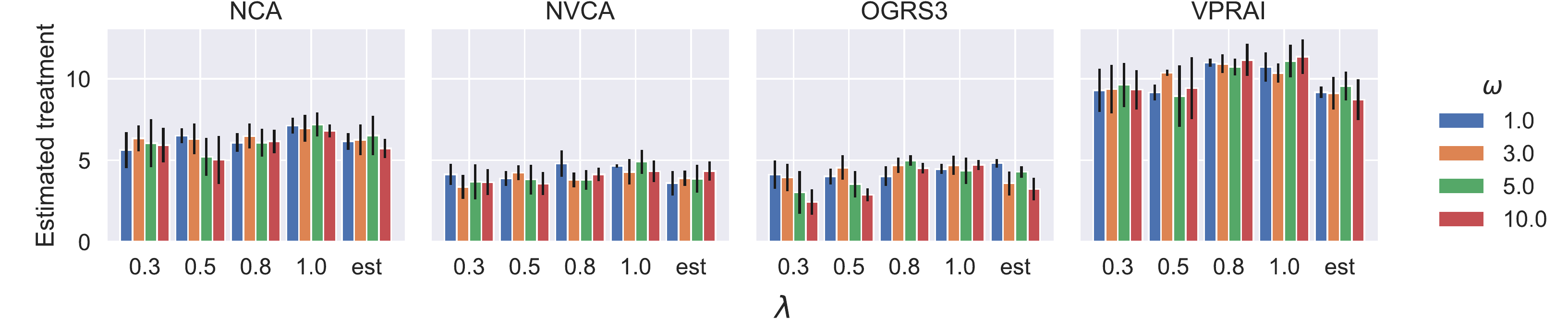}
    \caption{Bar plots of the results of the sensitivity analysis across different values of $\lambda$ and $\omega$. The $Y$ axis shows the Average Effect (AE), when comparing subpopulations of Black and White individuals for the NCA (a), NVCA (b),  OGRS3 (c), and VPRAI (d) scores. The difference in AE is presented as the percentage of the score range of the RAI. The value of $\lambda$ is shown on $X$ axis, and the value of $\omega$ is denoted by color. Errors are calculated as the standard deviation of five runs with different seeds. We can see that the magnitude of the effect is relatively stable to the change in parameters. All experiments presented ran with $\Delta T = 10$; matched on the largest set of crime bins, i.e., \{0, 1, 2, 3, 4--5, 5--6, 7--9, 10--19, 20--49, 50+\}; and used aggregated arrest rates (see \Cref{sect:arrest_rate_estim} for details).}
    \label{fig:sensitivity}
\end{figure*} 

\begin{figure*}[t]
    \centering
    \includegraphics[width=\linewidth]{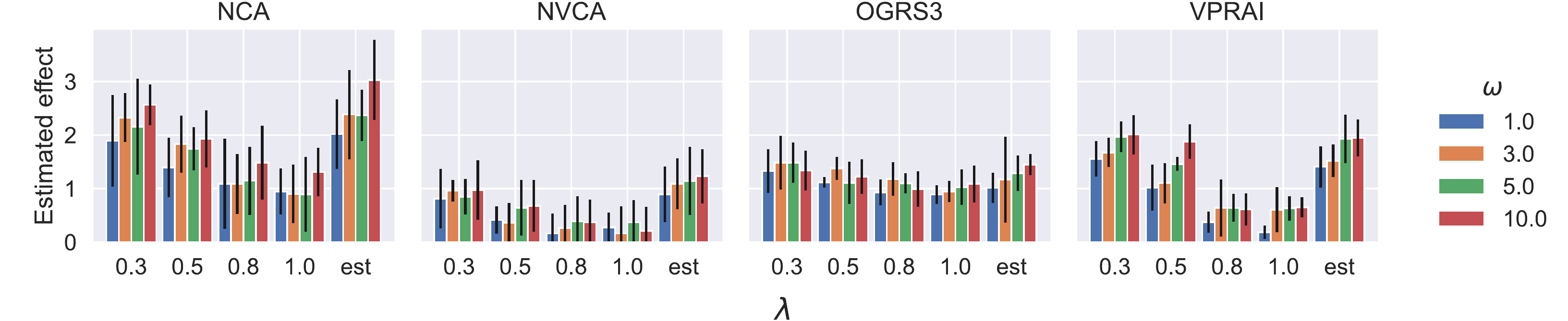}
    \caption{Bar plots of the results of the sensitivity analysis across different values of $\lambda$ and $\omega$ for the Average Effect (AE) when comparing Hispanic and White (non-White) subpopulations for the NCA (a), NVCA (b),  OGRS3 (c), and VPRAI (d) scores. The difference in AE is presented as the percentage of the score range of the RAI. The value of $\lambda$ is shown on $X$ axis, and the value of $\omega$ is denoted by color. Errors are calculated as the standard deviation of five runs with different seeds. With the exception of OGRS3 (c), we can see that the magnitude of the effect varies significantly with the parameters. All experiments presented ran with $\Delta T = 10$; matched on the largest set of crime bins, i.e., \{0, 1, 2, 3, 4--5, 5--6, 7--9, 10--19, 20--49, 50+\}; and used aggregated arrest rates (see \Cref{sect:arrest_rate_estim} for details).}
    \label{fig:sensitivity_Hisp}
\end{figure*}

\subsection{Sensitivity analysis}
\label{sec:results_sensitivity}

As discussed in \Cref{sect:crime_assignment}, the existing data have several limitations;
for example, the lack of Harris County arrest rate estimates forced us to use national level statistics instead.
We therefore need to examine how robust are our results (\Cref{tab:main_results}) with respect to a variety of plausible alternative modeling choices. 
We experiment with the following settings:
\begin{enumerate}[leftmargin=1.75em]
    \item \looseness=-1
    \textbf{Offense count bins} (\Cref{sect:estimating_effect}): 
    Bins \{0, 1, 2, 3--4, 5--6, 7--9\} are used in all experiments.
    We additionally vary the highest bins between \{10+\}, \{10--19, 20+\}, and \{10--19, 20--49, 50+\}.

    \item 
    \textbf{Arrest rate regularization} (\Cref{sect:arrest_rate_estim}):
    Namely \emph{averaging} and \emph{linear regression}.
    
    \item
    \textbf{Re-offense rate among the cohort} (\Cref{sect:lambda_estim}):
    We test two alternatives:
    (a)~$\lambda_g^c$ is set to a constant $\lambda \in \{0.3,0.5,0.8,1\}$ for all $c$ and $g$;
    (b)~$\lambda_g^c$ is estimated as described in \Cref{sect:lambda_estim}.
    
    \item
    \textbf{Influence of prior arrests on re-offending} (\Cref{sec:omega}):
    We vary $\omega \in \{1, 3, 5, 10\}$.  
    
    \item 
    \textbf{Temporal effect of prior convictions} (\Cref{sec:time_window}):
    We vary $\Delta T \in \{5, 10, 15, 20\}$.
\end{enumerate}

\looseness=-1
The results of the sensitivity analysis for varying $\lambda$ and $\omega$ are presented in \Cref{fig:sensitivity,fig:sensitivity_Hisp}.
We can see that the extent of the effect depends more on the RAI (e.g., NCA vs. NVCA\footnote{\looseness=-1 Unlike NCA, NVCA focuses only on \emph{violent} crime. The sensitivity to drugs \& DUI is a likely cause of NCA's bias, given the arrest disparities (\Cref{fig:arrest_rates_nsduh}).}) than on the choice of $\omega$ and $\lambda$. Generally speaking, estimating $\lambda$ (\Cref{sect:arrest_rate_estim}) results in slightly lower disparities. 
This is expected, as the estimated $\lambda$ (\Cref{apx:lambda}) are generally larger for the Black subpopulation, as an indirect result of higher arrest rates. 
This means more individuals in the White subpopulation 
with the same offending profile 
have never been arrested, 
and thus are not in the Neulaw cohort, excluding them from the comparison. 
We can interpret this as a `shift' in the disparities such that only a part of the selection bias due to differential arrest rates is captured when using the estimated values of $\lambda$. 

\looseness=-1
The sensitivity analysis for the Hispanic and White (non-Hispanic) subpopulations is in \Cref{fig:sensitivity_Hisp}. 
Coincidentally, the results presented in \Cref{tab:main_results} corresponds to the $\lambda$ values that produce the lowest effect for this population.
In contrast to the Black to White comparison above, the estimated values of $\lambda$ produce, generally speaking, the highest average effect. 
Overall, the results are much more sensitive to $\omega$ and $\lambda$.
This instability reinforces that the data limitations of NCVS may be too severe to obtain reliable estimates for the Hispanic subpopulations (see \Cref{sect:arrest_rate_estim} and the beginning of \Cref{sect:synth_comparison}).

\looseness=-1
The rest of this section is devoted to further investigation of the differences between RAI scores assigned to Black and White individuals.
Sensitivity analysis with respect to offense count bins, arrest rate regularization technique, and the restriction on temporal effect of prior convictions can be found in \Cref{apx:sensitivity}.
While minor variations are present, the results remain qualitatively similar to those presented in \Cref{tab:main_results}.

\subsection{Effect on subpopulations}

We now revisit the results presented in \Cref{tab:main_results}, and calculate AE, with respect to the arrest-only baseline (\Cref{tab:baseline}), for each intersection of gender and race. 
The results are shown in \Cref{tab:break}. 
We can see that for all scores, AE is larger for males, and, for all but the NCA, the older age group. 
The differences between males and females can be traced back to the arrests rates (\Cref{fig:arrest_rates_nsduh,fig:arrest_rates_ncvs}), where the disparities in arrests are generally more pronounced for men.
As for age, this may be because older individuals have had longer time to accumulate arrests which increases their probability of being assigned unobserved crimes (see \Cref{eq:crime_weight}).

\begin{table*}[ht]
\centering
\caption{Average percentage point difference in NCA, NVCA, VPRAI and OGRS3 scores between matched subpopulations.}
\label{tab:break}
\begin{tabular}{llcccc}
\toprule
Gender & Age & NCA & NVCA & OGRS3 & VPRAI  \\
\midrule
\multirow{2}{*}{Male} & 
18--29 &
$8.27\%$ $\pm1.48\%$ &
$3.12\%$ $\pm1.68\%$ & 
$3.00\%$ $\pm1.17\%$ &
$4.97\%$ $\pm1.33\%$ \\
    & 
30+ &
$7.09\%$ $\pm0.67\%$ &
$4.17\%$ $\pm0.65\%$ & 
$4.33\%$ $\pm0.46\%$ &
$9.11\%$ $\pm0.70\%$ \\
\midrule
\multirow{2}{*}{Female} & 
18--29 &
$1.74\%$ $\pm2.11\%$ &
$2.72\%$ $\pm3.51\%$ & 
$0.32\%$ $\pm1.39\%$ &
$2.90\%$ $\pm2.23\%$ \\ 
    & 
30+ &
$3.76\%$ $\pm1.02\%$ &
$3.89\%$ $\pm0.92\%$ & 
$3.24\%$ $\pm0.88\%$ &
$5.82\%$ $\pm1.07\%$ \\ 
\bottomrule
\multicolumn{6}{l}{\parbox[t][][t]{0.8\textwidth}{\vspace{0.3em}\footnotesize{Difference in Average Effect (AE) change from baseline in the average difference in risk scores between matched groups of Black and White individuals, when matching on age, sex, and simulated criminal history and arrest records. The baseline is the equivalent AE based on arrests records only. AEs are presented as a percentage of the score range of the respective RAI. 
Results are presented for a simulation with $\lambda=1$, $\omega=1$, $\Delta T=10$ and matched on the largest set of crime bins, i.e., \{0, 1, 2, 3, 4--5, 5--6, 7--9, 10--19, 20--49, 50+\}.}}}
\end{tabular}
\end{table*}

\subsection{What if the arrest rate discrepancies are even larger?}
\label{sect:arrest_disparities_higher}

\looseness=-1
As mentioned in \Cref{sect:arrest_rate_estim}, 
the arrest rate estimates we use come from national surveys,
whereas our analysis is constrained to the Harris County.
There is evidence that arrest rate disparities vary significantly with location, 
particularly for certain crime types like drug use and possession \citep{ButRobZil2022}. 
To measure sensitivity of the RAIs to variation in the arrest rate disparities, we conducted experiments with artificially increased disparities.\footnote{We only consider higher disparity to illustrate the potential level of harm. Lower disparity would not increase harm.} 
To achieve this, we multiply the estimated arrest rates 
for the Black subpopulation by a constant. 
The results are presented in \Cref{tab:mult}.
We can see that multiplying the arrest rate by five doubles the AEs relative to the unaltered results. 
The cause may be that as the difference in the amount of unobserved offenses grows, there are less individuals that can be matched. 
Although these results are highly artificial, they give us a sense of the sensitivity of RAI scores to changes in the arrests rates themselves.

\begin{table*}[h]
\centering
\caption{Average percentage point difference in NCA, NVCA, VPRAI and OGRS3 scores with synthetic arrest rates.}
\label{tab:mult}
\begin{tabular}{lcccc}
\midrule
\multicolumn{5}{c}{Black--White} \\
\midrule
Arrest rates & NCA & NVCA & OGRS3 & VPRAI  \\
\midrule
No change &
$7.12\%$ $\pm0.25\%$ &
$4.65\%$ $\pm0.06\%$ & 
$4.47\%$ $\pm0.16\%$ &
$10.72\%$ $\pm0.46\%$ \\
x2 for Black individuals  &
$8.48\%$ $\pm0.30\%$ &
$6.13\%$ $\pm0.41\%$ & 
$4.95\%$ $\pm0.20\%$ &
$11.27\%$ $\pm0.32\%$ \\

x3 for Black individuals &
$11.12\%$ $\pm0.33\%$ &
$7.65\%$ $\pm0.34\%$ & 
$6.59\%$ $\pm0.36\%$ &
$14.33\%$ $\pm0.75\%$ \\ 
x5 for Black individuals &
$14.20\%$ $\pm0.67\%$ &
$9.70\%$ $\pm0.97\%$ & 
$8.62\%$ $\pm0.72\%$ &
$18.22\%$ $\pm1.24\%$ \\ 
\hline
\multicolumn{5}{l}{\parbox[t][][t]{0.8\textwidth}{\vspace{0.3em}\footnotesize{Average Effect between matched groups of Black and White individuals, when matching on age, sex, and simulated criminal history and arrest records. The arrests rates for all subgroups containing Black individuals have been synthetically multiplied by 2, 3, and 5, respectively. AEs are presented as a percentage of the score range of the respective RAI. 
Results are presented for a simulation with $\lambda=1$, $\omega=1$, $\Delta T=10$ and matched on the largest set of crime bins, i.e., \{0, 1, 2, 3, 4--5, 5--6, 7--9, 10--19, 20--49, 50+\}.}}}
\end{tabular}
\end{table*}

\section{Discussion}
\label{sec:discussion}

\subsection{Limitations, ethics, and social impact}
\label{sec:ethics}

\looseness=-1
\subsubsection*{Data privacy}

\looseness=-1
Neulaw contains personal data of vulnerable individuals.
This data is publicly available, collected, and published by government officials and academics \citep{zilka2022a, ncvs,nsduh,ormachea2015new}. 
All data has been anonymized (by others), and we present results at the group level.

\subsubsection*{Data limitations}
\citet{knox2020administrative} highlight the difficulty of investigating racial bias when ``police administrative records lack information on civilians police observe but do not investigate''. We acknowledge this limitation and its impact on this analysis. Nonetheless, with the absence of more suitable data, we attempted to follow \citet{knox2020toward} by carefully considering counterfactuals and clearly stating our assumptions. We highlight further limitations below. 

\looseness=-1
We use arrest rates estimates from national self-reported surveys, conducted over the general population. 
This makes the estimates susceptible to several concerns: 
1)~reliability of self-reporting (under- or over-reporting of criminal activity, arrests, and victimization experiences); 
2)~exclusion of relevant populations (e.g., incarcerated or destitute people); and
3)~failure to adjust for regional arrest rate variations.
The latter is specifically problematic as all arrest records we examine are from the Harris County.
Since Harris County is, to our knowledge, the only public source containing unique individual identifiers (\Cref{sect:creating_cohort}),\footnote{The full COMPAS dataset contains unique individual identifiers as well, but it also contain full names, which we consider unethical to use here \citep{zilka2022a}.} which we need to compile arrest histories, we cannot analyze and compare results across locations.
We also note that we calculate the RAI score based on court records found in Neulaw, rather than strictly on arrest records (not all arrests lead to prosecution) or convictions (not all cases result in the defendant being found guilty). This may mean that our results are conservative, as the step from arrests to convictions is also likely to be impacted by racial disparities. 

\subsubsection*{Race and ethnicity recording} 

\looseness=-1
When combining information from three datasets, we can only use racial and ethic categories that appear in all datasets. This means that we exclude many racial and ethnic identities from our analysis. In addition, the way race and ethnicity are recorded over the datasets is inconsistent. 
In NSDUH, both race and ethnicity is self-reported.
On the contrary, in NCVS these are based on the victim's impression or knowledge. 
In Neulaw, it is not clear whether race is based on the individual’s self-description or not, and Hispanic decent is estimated by the database creator based on the individual's last name. 
Although this is not ideal, as the Harris County population has a large proportion of Hispanics ($\approx 40\%$) \citep{census2022quickfacts}, considering Hispanics as part of the White population group may cause a substantial error \citep{steffensmeier2011reassessing}. 
We recommend that high-quality datasets are curated, particularly regarding the Hispanic community, in order to investigate the robustness of our results.

\subsubsection*{Limitations of the methodology.} 

\looseness=-1
A key modeling simplification we make is in the assignment of unobserved crimes to members of the cohort. 
Both parameters used in our assignment model, $\lambda$ and $\omega$, cannot be measured from our data alone. As part of the sensitivity analysis, we attempt to estimate $\lambda_g^c$ from the data (see \Cref{sect:lambda_estim}). However, we do not vary $\omega$ by crime type and subgroup, even though this parameter is likely to vary in practice (see \Cref{sect:arrest_rate_estim} for more details).
Nonetheless, the effects we found persist throughout the sensitivity analysis. 
While we do not capture individual-level impact, our results provide strong evidence that RAIs can suffer from bias which cannot be detected using methods relying solely on arrests.  

\subsubsection*{Choice of RAIs.} 
We were only able to analyze open-source RAIs, i.e., RAIs that are fully transparent about the way scores are calculated (unlike, for example, COMPAS). 
However, this analysis is not possible if we cannot replicate the score algorithm, which is the case for most commercial RAIs.   




\subsubsection*{Societal impacts of RAIs and broader impacts of this research.}

There is wide literature covering concerns around the use of RAIs \citep[e.g.,][]{hamilton2020judicial,hamilton2015risk,monahan2016risk,mayson2019bias,eaglin2017constructing} ranging from their scientific validity to ethical concerns. 
RAIs are often introduced with the hope that they will eventually improve the objectivity and consistency of pre-trial and sentencing decisions. 
Several RAIs, including the ones presented in this paper, are completely transparent about the way the score is calculated, and VPRAI was even released alongside a publicly available dataset to allow for external validation.
However, the implications of using re-arrest as the primary training and validation objective are rarely addressed or acknowledged. 

The aim of this work is to add to this body of literature and improve our understanding of the impacts of RAIs, in order to inform policy and research concerning such tools, in service of a more equitable criminal justice system. In particular, we hope to raise awareness of the potential impact of differential reporting/discovery on these tools, and the follow-on effects to those subject to such tools. We also hope to inspire the research community to use quantitative methods to investigate these issues, alongside the vital qualitative work being undertaken, and to take the impact of differential reporting/discovery into account in related work. Such quantitative and qualitative investigation is crucial to inform our understanding on how (or whether) these tools should be developed and used.

Despite our intentions for positive impact, we acknowledge that our research poses societal risks. 
The use of racial and ethnic categories in this work risks reifying these constructs and contributing to harmful stereotypes \citep{benthall2019racial}. 
Scholars have also articulated the problematic nature of modeling race within causal modeling \citep{kohler2018eddie}. 
We do not intend to imply that race is an inherent property that is itself responsible for differences in behavior, criminality, or treatment. 
Concentrating on the accuracy of predictive tools further risks ignoring their other impacts, and wider concerns on how decisions are made within the justice system. 

\looseness=-1
Some of our simplifying assumptions may dampen the effect that we measure---for example, using estimates from national datasets, which likely mask large variations across regions. This may lead to a false sense of security about the true magnitude of the effect. More generally, the limitations of our approach mean that the numerical results are, of course, uncertain.
While we have tried to mitigate this through sensitivity analysis and being transparent about the limitations, the risk of under- or overstating the effect remains present.


\vspace{-1em}

\subsection{Conclusion}
\label{sect:conclusion}

\looseness=-1
We provide quantitative evidence that RAIs exhibit bias stemming from a failure to adjust for subpopulation level disparities in rates of arrest, crime reporting, and discovery of criminal activity.
While the average differences between scores for matched Black and White individuals range between 0.5--2.8\%, 
the gap grows to 4.4--11.0\% when we include estimated unobserved crime.
The size of the difference demonstrates that the use of arrest records as a proxy for crime biases not only the RAIs, but also their \emph{evaluation}.
This pattern is robust under our sensitivity analysis. 
Our results for Hispanic and White individuals are less reliable due to the discussed data limitations; further work is needed to verify these.

\looseness=-1
While quantifying the impact of arrest rate disparities on RAIs and their evaluation is complex (\Cref{sect:crime_assignment}),
that does not mean that we should ignore these effects.
In fact, the effects may be even more pronounced in practice since our use of national arrest rate averages could lead to an underestimate in certain areas (see \Cref{sect:arrest_disparities_higher}).
In addition, prior work warns that the deployment of predictive policing can increase disparities in arrest rates \citep{lum2016predict, sankin_2021_crime}.
This---in combination with our results---raises a concern that predictive policing tools may further impair the reliability of RAIs.

\looseness=-1
Continued reliance on RAIs without adjustment for the effect of disparate arrest rates lends a false impression of objectivity to potentially biased tools.
We therefore encourage researchers and developers to look beyond the observational data when evaluating RAIs and other decision tools deployed in criminal justice.
As algorithmic tools are unlikely to disappear, we need further research exploring how to account for the biases stemming from unobserved effects, including how observational data are often a poor proxy for the complicated reality of the criminal justice system.

\looseness=-1
In response to our findings, we recommend the following: 
\begin{enumerate}[leftmargin=1.75em]
    \item When designing RAIs, arrest rate disparities need to be taken into account.
    For example, this may translate into discounting arrests for offences 
    where the likelihood of arrest given a crime was committed is believed to vary significantly between subpopulations
    (e.g., drug possession).
    
    \item \looseness=-1
    RAI evaluation should be done at a sufficient level of geographic granularity, and incorporate an assessment of the impact of the local enforcement patterns, including variations in arrests rates across subpopulations and neighborhoods. 
    
    \item \looseness=-1
    Responsible development and evaluation of algorithmic tools necessitates reliable and accurate data. 
    In the short term, better access to local estimates will enable accounting for local variation and highlighting regions of particular concern. 
    In the long term, longitudinal data on offending and arrests offers a path towards understanding the relationship between RAIs assessment and underlying risk level.
\end{enumerate}

\begin{acks}
We thank the reviewers for their helpful comments and suggestions. M.Z. acknowledges support from the Leverhulme Trust grant ECF-2021-429. R.F. acknowledges support from PwC through Carnegie Mellon University's Digital Transformation and Innovation Center. A.W. acknowledges support from a Turing AI Fellowship under grant EP/V025279/1, The Alan Turing Institute, and the Leverhulme Trust via CFI.
\end{acks}


\bibliographystyle{ACM-Reference-Format}
\bibliography{pipeline}

\newpage
\appendix
\beginsupplement

\onecolumn

\section{How we calculated the RAI scores}
\label{Apx:scores}

\subsection{NCA}

The NCA points are calculated as following \citep{NCA}:

\begin{itemize}
    \item Age at current arrest: 2 points if 22 or younger, 0 otherwise. 
    \item Pending charges at the time of the arrest: 3 points if yes, 0 otherwise. Calculated by checking if the dataset contains a prior arrest that have not been disposed before the date of the current arrest. 
    \item Prior misdemeanor conviction: 1 point if yes, 0 otherwise. Calculated by checking if the dataset contains a prior arrest for a misdemeanor that was disposed as "guilty".
    \item Prior felony conviction: 1 point if yes, 0 otherwise. Calculated as above, but for felonies instead of misdemeanors.
    \item Prior violent conviction: 2 points if 3 convictions or more, 1 point if 1 or 2 convictions, 0 otherwise. Calculated as above, but only for violent offenses. We used the definition for "violent offense" found in the Texas Code of Criminal Procedure, article 17.50,  \citep{vcrime}.
    \item Prior failure to appear in the past 2 years: 2 points if 2 or more, 1 point if 1, 0 otherwise. Calculated by checking if the dataset contains a prior arrest from the past 2 years for a FTA charge. 
    \item Prior sentence to incarceration: 2 points if yes, 0 otherwise. Calculated by checking if the dataset contains a prior arrest that was disposed in one of the following ways: committed to local jail, committed to TDC, state jail, life sentence or shock probation. 
    
\end{itemize}

The score is then calculated according to Table~\ref{tab:nca_score}. 
\begin{table}[h]
    \centering
    \caption{NCA score calculation}
    \begin{tabular}{cc}
    Total NCA points & NCA score \\
    \midrule
    0 & 1 \\
    1 or 2 & 2 \\
    3 or 4 & 3 \\
    5 or 6 & 4 \\
    7 or 8 & 5 \\
    9 to 13 & 6 \\
    \end{tabular}
    \label{tab:nca_score}
\end{table}

\subsection{NVCA}

The NVCA points are calculated as following \citep{NCA}:

\begin{itemize}
    \item Current violent offense: 2 points if yes, 0 otherwise. Calculated based on the definition for "violent offense" found in the Texas Code of Criminal Procedure, article 17.50 \citep{vcrime}. 
    \item Current violent offense and 20 year old or younger: 1 point if yes, 0 if otherwise.
    \item Pending charges at the time of the arrest: 1 points if yes, 0 otherwise. Calculated by checking if the dataset contains a prior arrest that have not been disposed before the date of the current arrest. 
    \item Prior conviction: 1 point if yes, 0 otherwise. Calculated by checking if the dataset contains a prior arrest that was disposed as "guilty". 
    \item Prior violent conviction: 2 points if 3 convictions or more, 1 point if 1 or 2 convictions, 0 otherwise. Calculated as above, but only for violent offenses. We used the definition found in the Texas Code of Criminal Procedure, article 17.50, for "violent offense" \citep{vcrime}.
    
\end{itemize}
The score is then calculated according to Table~\ref{tab:nvca_score}. 
\begin{table}[]
    \centering
    \caption{NVCA score calculation}
    \begin{tabular}{cc}
    Total NVCA points & NVCA score \\
    \midrule
    0 or 1 & 1 \\
    2 & 2 \\
    3 & 3 \\
    4 & 4 \\
    4 & 5 \\
    6 or 7 & 6 \\
    \end{tabular}
    \label{tab:nvca_score}
\end{table}

\subsection{VPRAI}

The score is then calculated as following \citep{VPRAI}: 
\begin{itemize}
    \item Charge type: 1 point if current arrest is a felony, 0 otherwise. 
    \item Pending charges at the time of the arrest: 1 points if yes, 0 otherwise. Calculated by checking if the dataset contains a prior arrest that have not been disposed before the date of the current arrest. 
    \item Outstanding Warrants: 1 points if yes, 0 otherwise. We were not able to calculate this factor. 
    \item Prior conviction: 1 point if yes, 0 otherwise. Calculated by checking if the dataset contains a prior arrest that was disposed as "guilty". 
    \item Prior violent conviction: 1 points if 2 convictions or more, 0 otherwise. Calculated as above, but only for violent offenses. We used the definition for "violent offense" found in the Texas Code of Criminal Procedure, article 17.50,  \citep{vcrime}.
    \item Prior failure to appear convictions: 2 points if 2 or more, 0 otherwise. Calculated by checking if the dataset contains a prior arrests for a FTA charges. 
    \item :ength of residence: 1 points if lived at current residence less than a year, 0 otherwise. We were not able to calculate this factor. 
    \item Employed or primary child caregiver: 1 points if no, 0 otherwise. We were not able to calculate this factor. 
    \item History of drug abuse: 1 point if yes, 0 otherwise. Calculated based on prior drugs convictions. 
\end{itemize}

    We note there is an updated version of the VPRAI that assigns different weights to the risk factors.

\subsection{OGRS3}

The score is calculated in a similar fashion the scores above, according to the coefficients found here: \citep{OGRS3}. 
The only adjustment is that we used arrests that did not lead to convictions instead of "caution/reprimand/warning".

\section{Arrest rates}

\begin{figure}[h]
    \centering
    \includegraphics[width=0.7\linewidth]{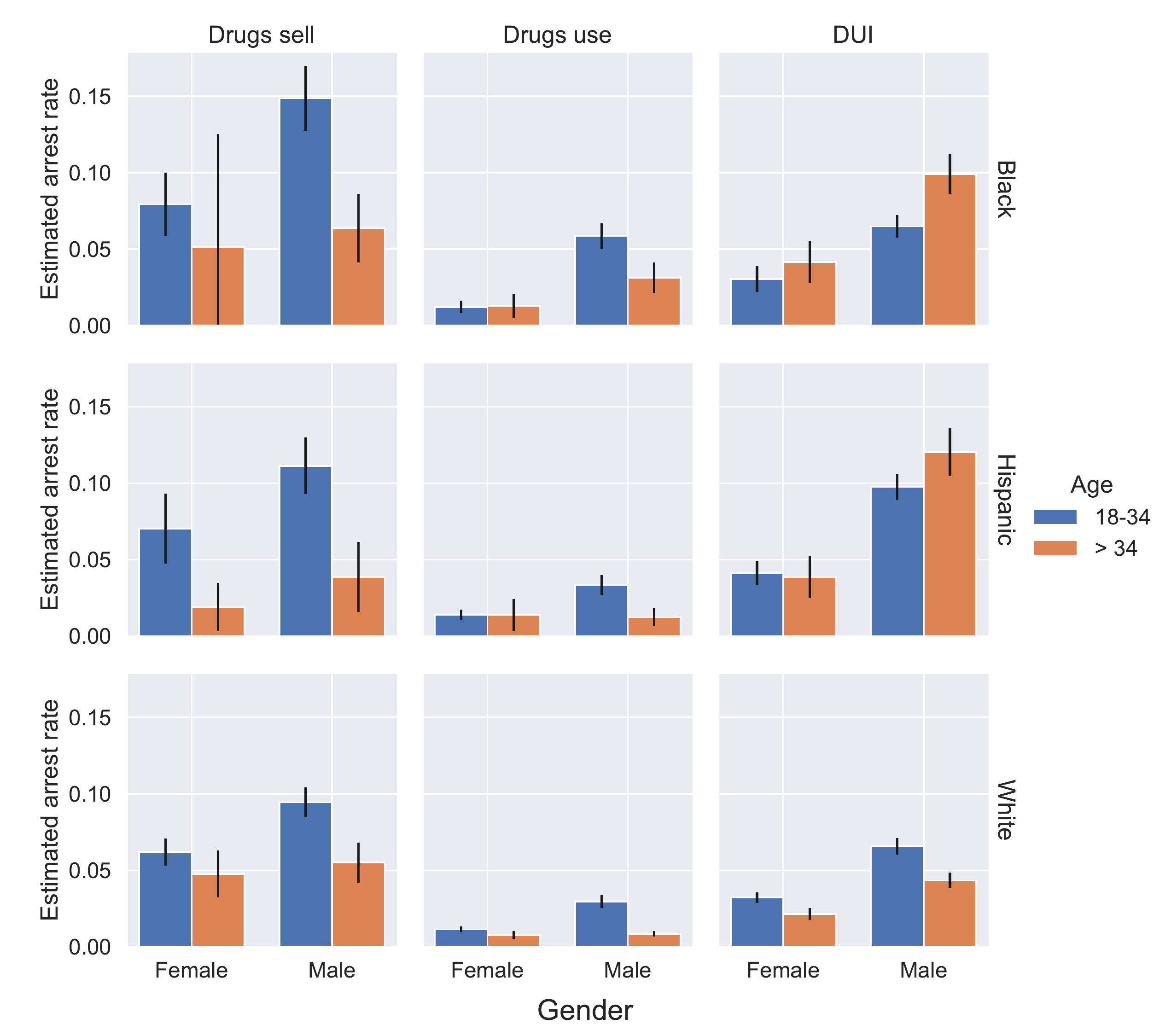}
    \caption{Average Estimates of arrest rates using a linear regression model by sex, age and race of people who committed drug use, drug sale, and DUI offenses. Rates are estimated from the National Survey on Drug Use and Health (NSDUH) as the average arrest rate for respondents that self-reported engaging in these illegal activities in the year prior to the interview.}
\label{apx:lr_arrest_rates_nsduh}
\end{figure}

\begin{figure}[h]
    \centering
    \includegraphics[width=\linewidth]{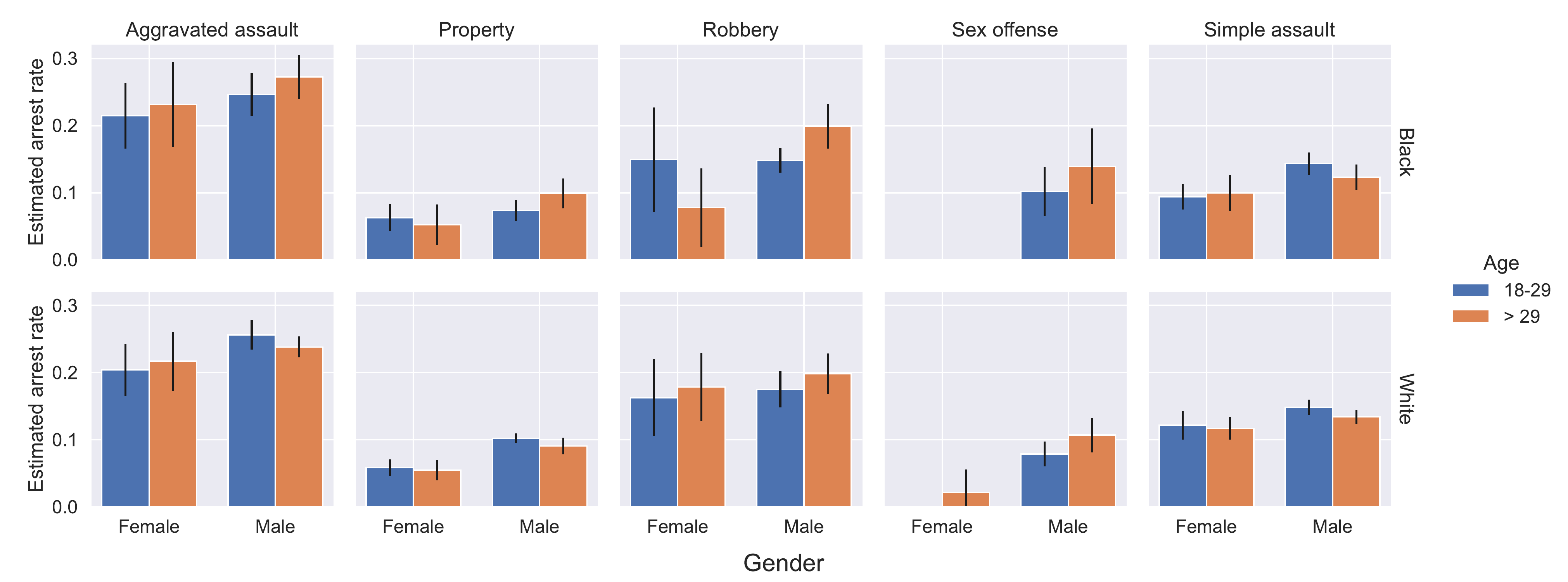}
    \caption{Average Estimates of arrest rates using a linear regression model by sex, age and race of people who committed property and violent offenses. Rates are estimated from the National Crime Victimization Survey (NCVS) as the average arrest rates for assaults, robbery, property and sexual offenses conditional on the  demographics of the person who committed the offense.}
    \label{apx:lr_arrest_rates_ncvs} 
\end{figure}

Arrest rates estimates based on a linear regression model. For rates calculated from NSDUH see \Cref{apx:lr_arrest_rates_nsduh}, and from NCVS see \Cref{apx:lr_arrest_rates_ncvs}.

\section{Calculated values of $\lambda$}
\label{apx:lambda}

See \Cref{tab:lamb_nsduh} for values for DUI and drug offenses calculated from NSDUH, and \Cref{tab:lamb_neulaw} for all other offense, calculated from Neulaw. For full details please see \cref{sect:lambda_estim}.

\begin{table}[]
\caption{$\lambda$ values calculated from the NSDUH dataset. }
\begin{tabular}{lllccc}
Race     & Sex    & Age             & DUI  & Drug sell & Drug use
\\
\hline
Black    & Female & 18-34           & 0.22 & 0.35      & 0.21     \\
Black    & Female & \textgreater 34 & 0.23 & 0.23      & 0.30     \\
Hispanic & Female & 18-34           & 0.21 & 0.40      & 0.19     \\
Hispanic & Female & \textgreater 34 & 0.21 & 0.13      & 0.24     \\
White    & Female & 18-34           & 0.16 & 0.29      & 0.17     \\
White    & Female & \textgreater 34 & 0.18 & 0.33      & 0.27     \\
Black    & Male   & 18-34           & 0.45 & 0.54      & 0.41     \\
Black    & Male   & \textgreater 34 & 0.51 & 0.53      & 0.52     \\
Hispanic & Male   & 18-34           & 0.39 & 0.49      & 0.38     \\
Hispanic & Male   & \textgreater 34 & 0.40 & 0.38      & 0.44     \\
White    & Male   & 18-34           & 0.37 & 0.47      & 0.37     \\
White    & Male   & \textgreater 34 & 0.38 & 0.53      & 0.48  \\
\midrule
\label{tab:lamb_nsduh}
\end{tabular}
\end{table}

\begin{table}[]
\caption{$\lambda$ values calculated from the Neulaw dataset. }
\begin{tabular}{lllccccc}
Race     & Sex    & Age             & Aggravated assault & Property & Simple assault & Sex offense & Robbery \\
\hline
Black    & Female & 18-29           & 0.52               & 0.33     & 0.46           & 0.53        & 0.63    \\
Black    & Female & \textgreater 29 & 0.55               & 0.41     & 0.54           & 0.61        & 0.75    \\
Black    & Male   & 18-29           & 0.81               & 0.67     & 0.71           & 0.66        & 0.76    \\
Black    & Male   & \textgreater 29 & 0.74               & 0.71     & 0.69           & 0.71        & 0.85    \\
Hispanic & Female & 18-29           & 0.40               & 0.23     & 0.35           & 0.41        & 0.54    \\
Hispanic & Female & \textgreater 29 & 0.47               & 0.25     & 0.37           & 0.49        & 0.65    \\
Hispanic & Male   & 18-29           & 0.68               & 0.56     & 0.56           & 0.48        & 0.70    \\
Hispanic & Male   & \textgreater 29 & 0.66               & 0.56     & 0.54           & 0.54        & 0.76    \\
White    & Female & 18-29           & 0.54               & 0.29     & 0.40           & 0.38        & 0.58    \\
White    & Female & \textgreater 29 & 0.48               & 0.35     & 0.46           & 0.48        & 0.69    \\
White    & Male   & 18-29           & 0.65               & 0.52     & 0.54           & 0.49        & 0.72    \\
White    & Male   & \textgreater 29 & 0.62               & 0.54     & 0.54           & 0.55        & 0.76  \\
\midrule
\label{tab:lamb_neulaw}
\end{tabular}
\end{table}

\section{Sensitivity analysis}
\label{apx:sensitivity}

\subsection{Varying the time-window}

\subsubsection{$\Delta T = 5$}
\begin{figure}[h]
    \centering
    \includegraphics[width=\linewidth]{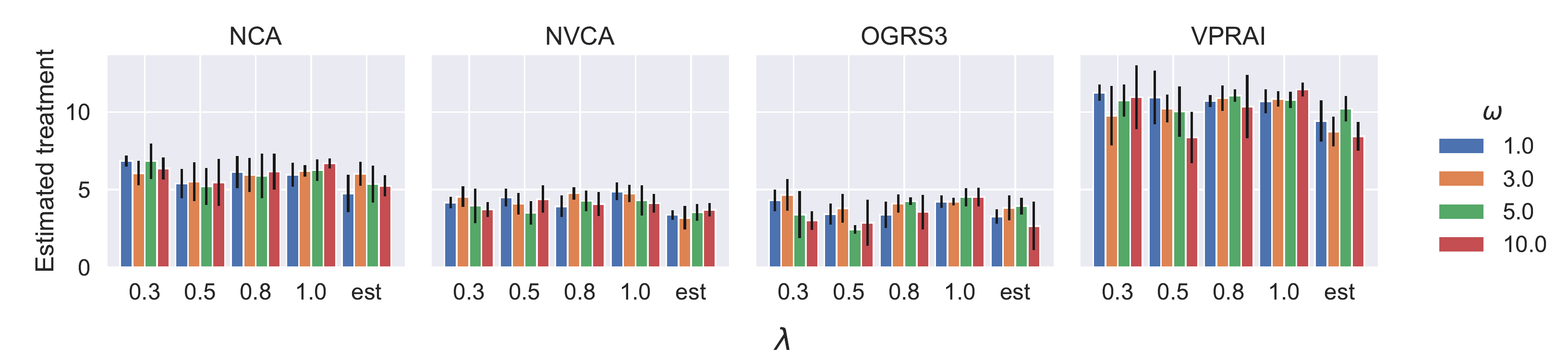}
    \caption{Bar plots of the results of the sensitivity analysis across different values of $\lambda$ and $\omega$. The $Y$ axis shows the Average Effect (AE), when comparing subpopulations of Black and White individuals for the NCA (a), NVCA (b),  OGRS3 (c), and VPRAI (d) scores. The difference in AE is presented as the percentage of the score range of the RAI. The value of $\lambda$ is shown on $X$ axis, and the value of $\omega$ is denoted by color. Errors are calculated as the standard deviation of five runs with different seeds. We can see that the magnitude of the effect is relatively stable to the change in parameters. All experiments presented ran with $\Delta T = 5$; matched on the largest set of crime bins, i.e., $\{0,1,2,3,4-5,5-6,7-9,10-19,20-49,50+\}$; and used aggregated arrest rates (see \Cref{sect:arrest_rate_estim} for details).}
    \label{fig:sensitivity_T_5}
\end{figure}

See \Cref{fig:sensitivity_T_5}.

\subsubsection{$\Delta T = 15$}
\begin{figure}[h]
    \centering
    \includegraphics[width=\linewidth]{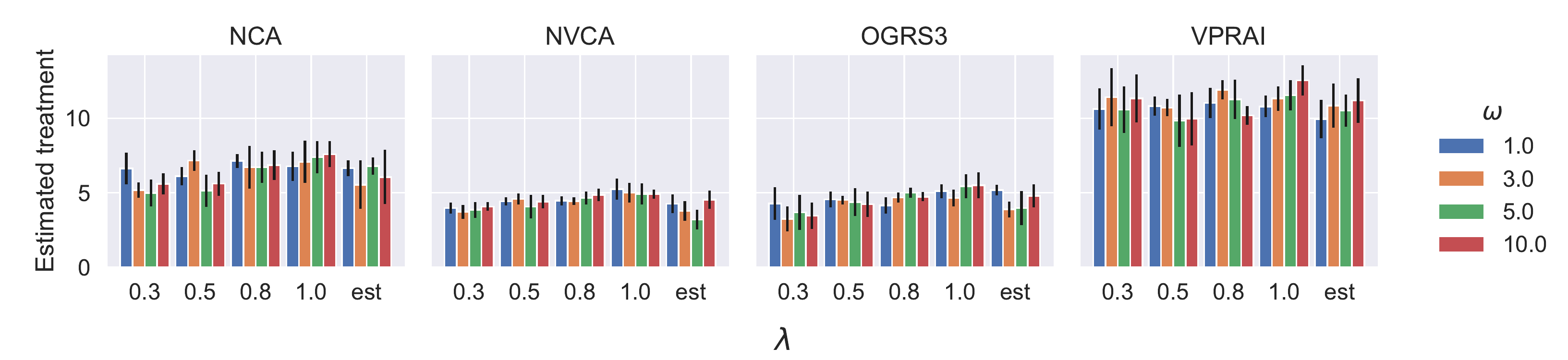}
    \caption{All experiments presented ran with $\Delta T = 15$; matched on the largest set of crime bins, i.e., $\{0,1,2,3,4-5,5-6,7-9,10-19,20-49,50+\}$; and used aggregated arrest rates.}
    \label{fig:sensitivity_T_15}
\end{figure}

See \Cref{fig:sensitivity_T_15}.

\subsubsection{$\Delta T = 20$}
\begin{figure}[h]
    \centering
    \includegraphics[width=\linewidth]{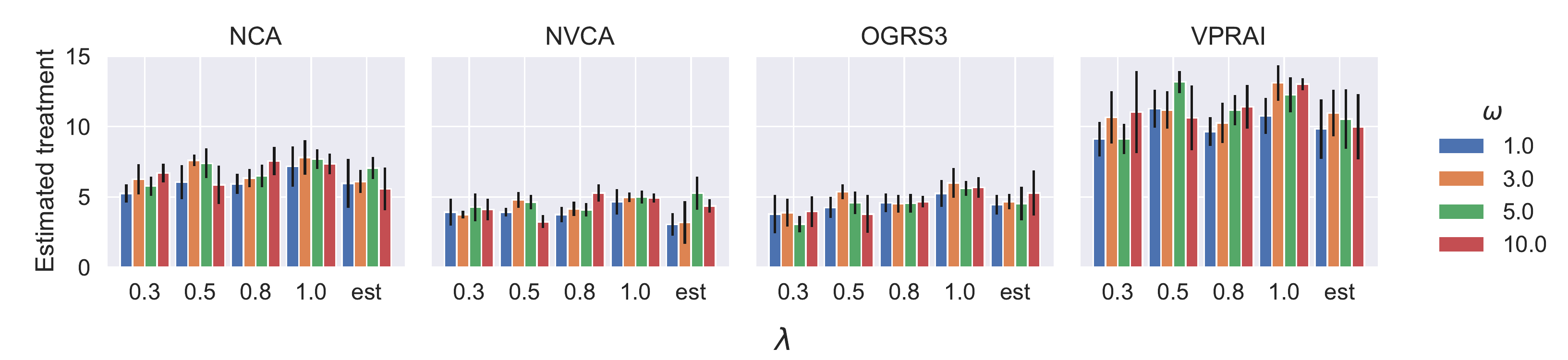}
    \caption{All experiments presented ran with $\Delta T = 20$; matched on the largest set of crime bins, i.e., $\{0,1,2,3,4-5,5-6,7-9,10-19,20-49,50+\}$; and used aggregated arrest rates.}
    \label{fig:sensitivity_T_20}
\end{figure}

See \Cref{fig:sensitivity_T_20}.

\subsection{Crime grouping}

\begin{figure}[h]
    \centering
    \includegraphics[width=\linewidth]{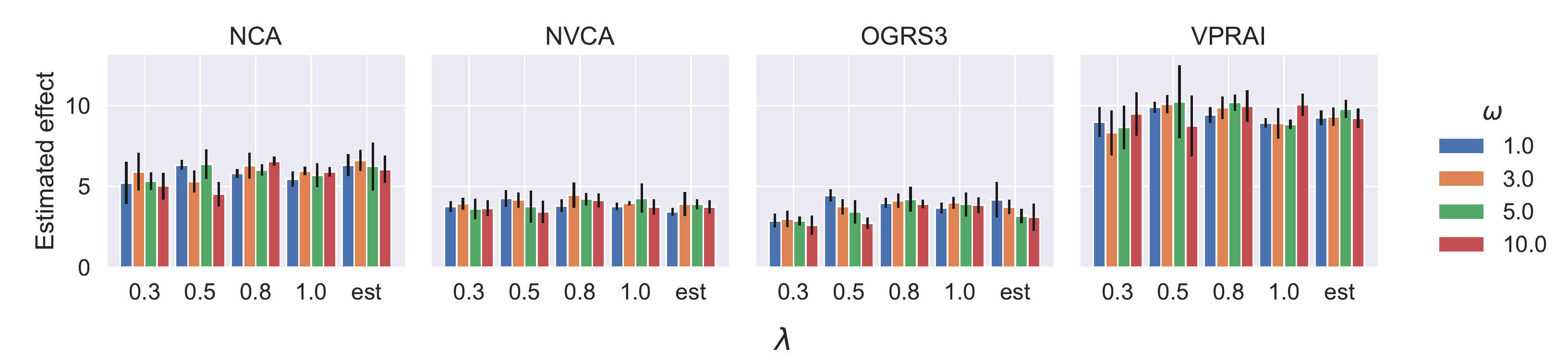}
    \caption{All experiments presented ran with $\Delta T = 10$; are matched on the following crime bins: $\{0,1,2,3,4-5,5-6,7-9,10-19,20+\}$; and used aggregated arrest rates.}
    \label{fig:sensitivity_B_20}
\end{figure}

\begin{figure}[h]
    \centering
    \includegraphics[width=\linewidth]{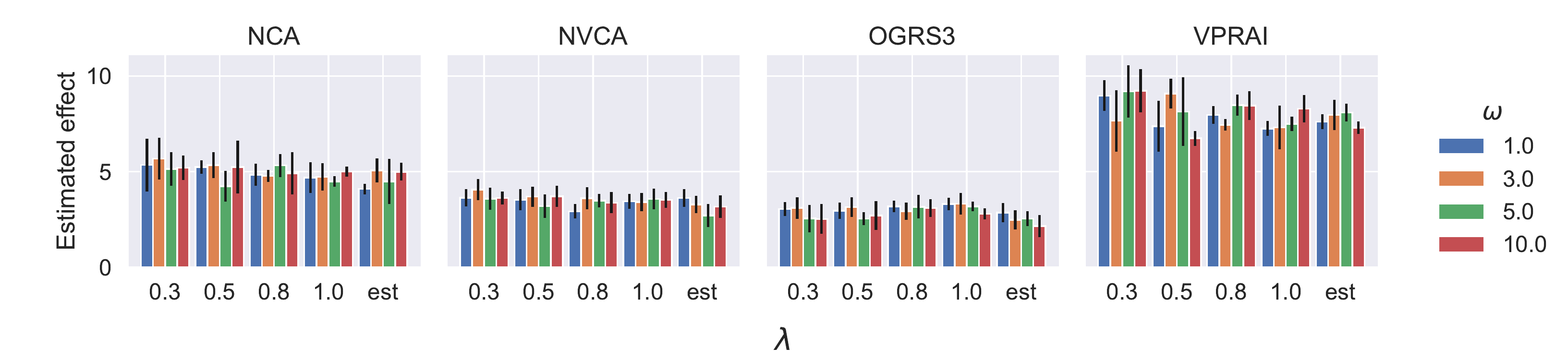}
    \caption{All experiments presented ran with $\Delta T = 10$; are matched on the following crime bins: $\{0,1,2,3,4-5,5-6,7-9,10+\}$; and used aggregated arrest rates.}
    \label{fig:sensitivity_B_10}
\end{figure}

See \Cref{fig:sensitivity_B_20} and \Cref{fig:sensitivity_B_10}.

\subsection{Method of calculating arrests rates}

\begin{figure}[h]
    \centering
    \includegraphics[width=\linewidth]{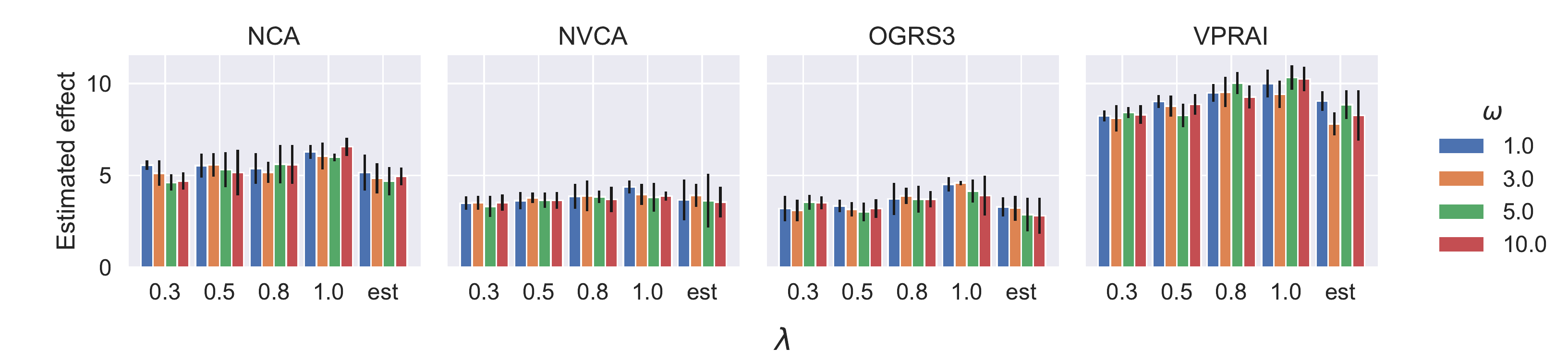}
    \caption{ All experiments presented ran with $\Delta T = 10$; matched on the largest set of crime bins, i.e., $\{0,1,2,3,4-5,5-6,7-9,10-19,20-49,50+\}$; and used the linear regression  arrest rates estimation method (see \Cref{sect:arrest_rate_estim} for details).}
    \label{fig:sensitivity_A_lr}
\end{figure}

See \Cref{fig:sensitivity_A_lr}.

\end{document}